\documentclass[aps,prl,twocolumn,superscriptaddress]{revtex4-1}
\usepackage{graphicx}
\usepackage{amssymb}

\newcommand{\dgr}{\,$^{\rm o}$}

\begin{document}

\title{Physical properties of FeSe$_{0.5}$Te$_{0.5}$ single crystals grown under different conditions}

\author{V. Tsurkan}

\affiliation{Experimental Physics 5, Center for Electronic Correlations and Magnetism, Institute of Physics, University of Augsburg, D 86159, Augsburg, Germany}

\affiliation{Institute of Applied Physics, Academy of Sciences of Moldova, MD 2028, Chisinau, R. Moldova}

\author{J. Deisenhofer}
\author{A. G\"unther}
\author{Ch. Kant}
\author{H.-A. Krug von Nidda}
\author{F. Schrettle}
\author{A. Loidl}

\affiliation{Experimental Physics 5, Center for Electronic Correlations and Magnetism, Institute of Physics, University of Augsburg, D 86159, Augsburg, Germany}

\date{[Received: date / Revised version: date ]}

\begin{abstract}

We report on structural, magnetic, conductivity, and thermodynamic studies of FeSe$_{0.5}$Te$_{0.5}$ single crystals grown by self-flux and Bridgman methods. The lowest values of the susceptibility in the
normal state, the highest transition temperature $T_c$ of 14.5~K, and the largest heat-capacity anomaly at $T_c$ were obtained for pure (oxygen-free) samples. The critical
current density $j_c$ of $8.6 \times 10^4$~A/cm$^2$ (at 2~K) achieved in pure samples is attributed to intrinsic inhomogeneity due to disorder at the anion sites. The samples containing spurious phase of Fe$_3$O$_4$ show increased $j_c$ up to $2.3 \times 10^5$~A/cm$^2$ due to additional pinning centers. The upper critical field $H_{c2}$ of $\sim 500$~kOe is estimated from the resistivity study
in magnetic fields parallel to the \emph{c}-axis. The anisotropy of the upper critical field $\gamma_{H_{c2}} = H_{_{c2}}^{ab}/H_{_{c2}}^{c}$ reaches a value $\sim 6$ at $T\longrightarrow T_c$. Extremely low values of the residual
Sommerfeld coefficient for pure samples indicate a high volume fraction of the superconducting phase (up to 97 \%). The electronic contribution to the specific heat in the
superconducting state is well described within a single-band BCS model with a temperature dependent gap $\Delta_0 = 27(1)$~K. A broad cusp-like anomaly in the electronic specific
heat of samples with suppressed bulk superconductivity is ascribed to a splitting of the ground state of the Fe$^{2+}$ ions at 2c sites. This contribution is fully suppressed in the ordered state in samples with bulk superconductivity.

\end{abstract}


\maketitle

\section{Introduction}

The recent finding of high-temperature superconductivity in iron pnictides \cite{KWH08,RTJ08,TTL08} and iron chalcogenides \cite{HLY08} inspired an immense research activity in complex materials
similar to the discovery of high-$T_c$ cuprate superconductors two decades ago. Among the new iron-based superconductors, the iron chalcogenides are of particular
interest due to a simple crystal structure (Fig. 1). It consists of Fe ions tetrahedrally coordinated by Se and Te arranged in layers stacked along the \emph{c}-axis in the
tetragonal lattice without any other interlayer cations, as for example in pnictides. Therefore, the iron chalcogenides were believed to be most suitable for
investigation of the interplay of competing orders and pairing mechanisms. However, recently it was recognized in Fe(Se,Te) that Fe ions can occupy also the 2c positions in the anion plane \cite{LD09,LCH09}.

\begin{figure}[t]
\centering
\includegraphics[angle=0,width=0.30\textwidth]{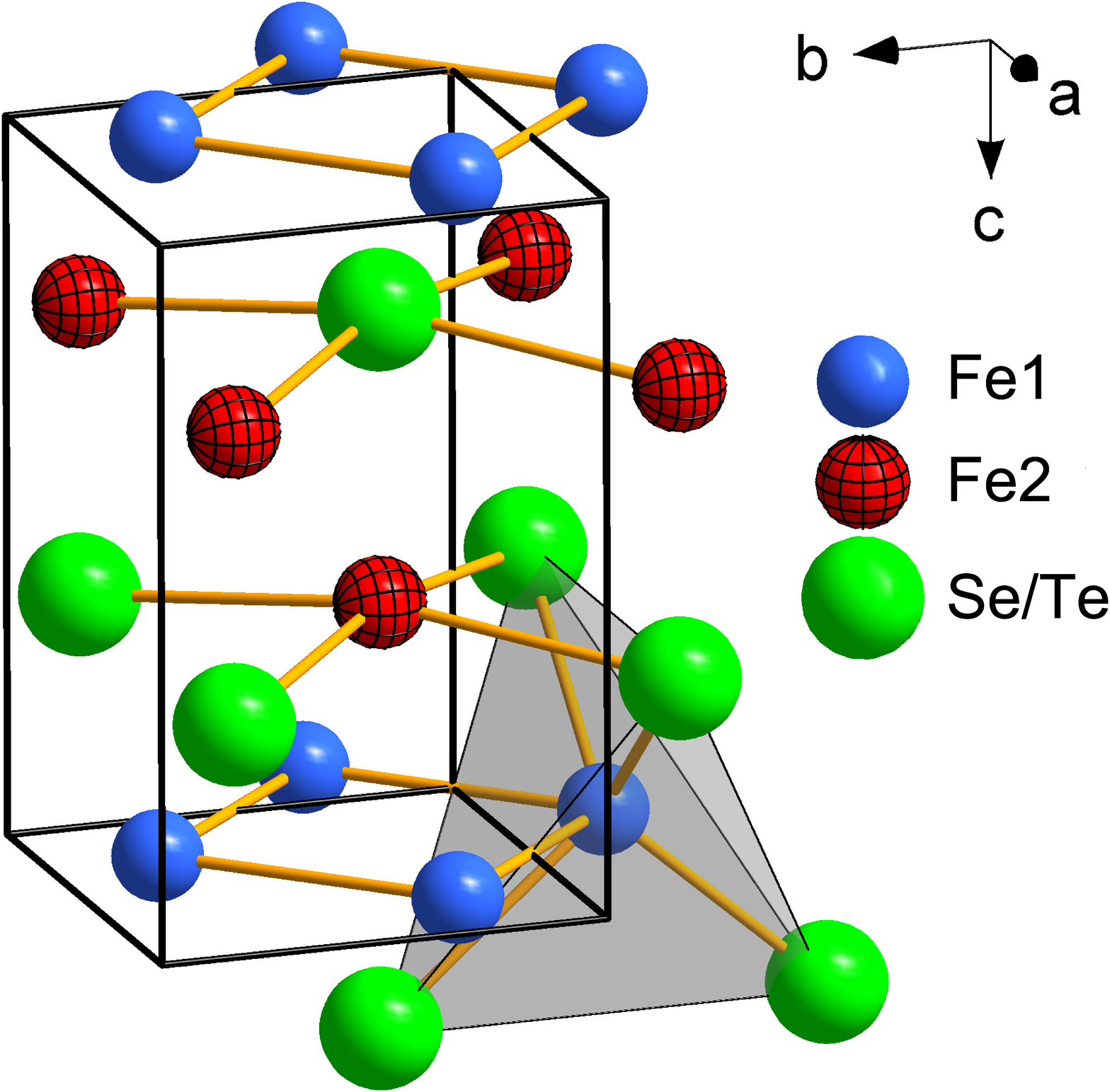}
 \caption{(color online) Crystal structure of FeSe$_{0.5}$Te$_{0.5}$ demonstrating
the tetrahedral coordination of Fe1 ions in 2a positions (blue spheres) by anions of Se and Te (green spheres)
 and the interstitial Fe2 ions in 2c position (red spheres)
in the anion plane.}
\end{figure}

Nearly stoichiometric FeSe becomes superconducting below 8 K at ambient pressure \cite{HLY08} but the transition temperature can be
enhanced up to 37 K by application of external pressure \cite{MMT09,MTO09}. The superconducting properties of FeSe were shown to be extremely sensitive to deviations from the
stoichiometry \cite{MHK09,PCP09}. The substitution of Se by Te in FeSe increases the temperature of the superconducting transition to a maximum of about 14 K for 50 \% of
replacement \cite{FPQ08,YHH08}. The intrinsic disorder due to random distribution of the Se and Te ions among the anion sites in the crystal lattice is expected to reduce the
sensitivity of FeSe to non-stoichiometry. Subsequently, several authors recently reported on the successful preparation of high quality single crystals of FeSe$_{1-x}$Te$_x$
using Bridgman \cite{SSM09,CCD09,KBA09} and flux methods \cite{TTN09,HBW09}. The existing data show large variations in magnetic and conducting behavior of Fe-Se(Te) obviously related to differences in
the growth conditions and purity of the starting materials. Hence, only little information about the intrinsic properties of  FeSe$_{0.5}$Te$_{0.5}$ is available so far.

Here we report on the magnetic susceptibility, resistivity and heat-capacity measurements of FeSe$_{1-x}$Te$_x$ with nominal concentration $x = 0.5$. The samples were prepared in
various ways and exhibit significantly different behavior which was investigated in the temperature range 2 -- 400~K and in external magnetic fields up to 140~kOe.

\section{Experimental}

Single crystals of FeSe$_{0.5}$Te$_{0.5}$ were grown by self-flux and Bridgman methods. Flakes and shots of high-purity elements, 99.99 \% Fe, 99.999 \% Se, and 99. 999 \% Te, were used in the
growth experiments. To get a minimal amount of oxide impurity we additionally purified Se (for growth runs F216) and Te (for growth runs F213 and F216) by zone melting. Handling of these samples was done in an argon box with residual oxygen and water content less than 1 ppm. The
growth of single crystals was performed in evacuated double quartz ampoules. For different runs in the self-flux method the soaking temperature varied between 900 and 1100\dgr\,C. The cooling rate varied between 1 and 60\dgr\,C/h. In the Bridgman method we used pulling rates between 0.5 and 2 mm/h and rotation speed of 2-5 turns/min. Final treatment of the samples in both preparation methods was done at 410\dgr\,C for 70 -- 100 hours followed by quenching in ice water. The plate-like samples with shiny faces with dimensions $\sim 3\times 5$~mm$^{2}$ in the \emph{ab} plane and thickness up to 0.5~mm along the \emph{c} axis were obtained on the top of the solidified boule. The single-crystallinity of the grown samples was checked by the single crystal X-ray diffraction.
The sample composition was investigated by Energy Dispersive x-ray (EDX) analysis. The phase content of the samples was analyzed by x-ray powder
diffraction (Cu K$_\alpha$ radiation, $\lambda = 1.540560$~{\AA}) on crushed single crystals using a STADI-P powder diffractometer (STOE \& CIE) with a position sensitive detector.

The magnetic measurements were performed in a temperature range 2 -- 400 K and in magnetic fields up to 50~kOe using a SQUID magnetometer MPMS~5 (Quantum Design). Electron-spin resonance (ESR) measurements were performed in a
Bruker ELEXSYS E500 CW spectrometer at X-band frequencies
(9.36 GHz) equipped with a continuous He
gas-flow cryostat in the temperature region 4.2 $\leq T\leq $ 300~K.
The heat
capacity was measured by a relaxation method using a Quantum Design physical properties measurement system (PPMS) in a temperature range 1.8 -- 300~K and magnetic fields
up to 90~kOe. The magnetic field was applied parallel to the \emph{c}-axis of the samples. The resistivity studies were done on rectangular samples by a four-point method in the
temperature range 2 -- 300~K and in magnetic fields up to 140~kOe using a He-flow cryostat (Oxford Instruments) and using the resistivity measurement option of the PPMS in
magnetic fields up to 90~kOe. The contacts were made by conductive silver paint.

\section{Experimental Results and Discussion}

\subsection{X-ray analysis}

\begin{figure}[htb]
\includegraphics[angle=0,width=0.5\textwidth]{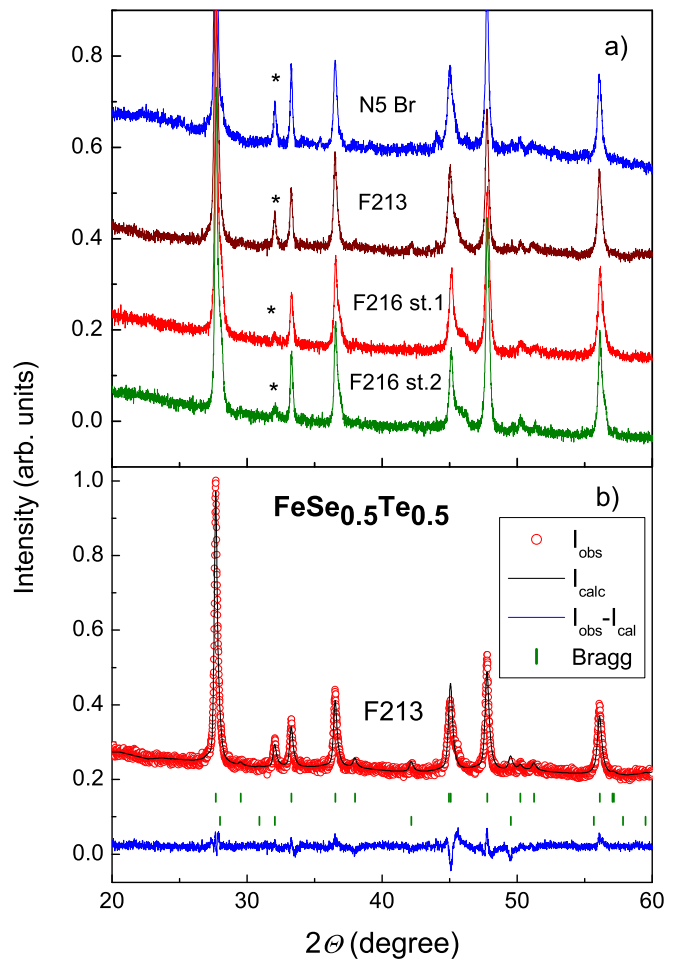}
 \caption{(color online) a) Room temperature x-ray diffraction patterns for FeSe$_{0.5}$Te$_{0.5}$ samples prepared under different conditions. Stars indicate the impurity phase; b) Experimental (open circles) and
refined (black line) x-ray diffraction patterns for sample F213. Blue line- difference between experimental and calculated
intensities. Vertical lines mark Bragg positions: top row- main tetragonal phase; bottom- impurity hexagonal phase.
Growth conditions: Sample BrN5 (Bridgman): soaking temperature T$_s$=1084\dgr\,C; soaking time t$_s$=24 h; puling rate 0.5 mm/h. Sample F213 (self-flux): T$_s$=1100\dgr\,C; t$_s$=72 h; cooling rate 60\dgr\,C/h. Sample F216 st. 1 (self-flux): T$_s$=1100\dgr\,C; t$_s$=72 h; cooling rate 60\dgr\,C/h. Sample F216 st. 2 (self-flux): T$_s$
=1100\dgr\,C; t$_s$=72 h; cooling rate 1\dgr\,C/h.}
\end{figure}

The x-ray diffraction patterns for the samples under investigation (Fig. 2a) are consistent with tetragonal symmetry P4/nmm for the main FeSe$_{0.5}$Te$_{0.5}$ phase  and with
hexagonal symmetry P63/mmc for the impurity phase Fe$_7$Se$_8$. The refinement was performed using the FULLPROF SUITE \cite{RC93}. The results of the refinement for one of the samples
are shown in Fig. 2b. The occupation of Te and Se at the 2c sites was refined constraining their sum to unity in agreement with the EDX data. A similar constraint was used for the occupation of Fe ions for the main phase allowing for two different sites (2a and 2c). The EDX measurements on samples from various growth experiments find no indication that the different methods used for the crystal growth yield different compositions. No significant gradient in the Te/Se ratio along the sample was observed. The Se content for samples from different growth runs varied from 0.455(6) to 0.498(10), and the respective variations of the Te content were within the range 0.555(4) to 0.502(10). The observed maximal deviations of Se/Te ratio for some samples of about 10~\% from the nominal 50/50~\% composition, correspond to  FeSe$_{0.45}$Te$_{0.55}$. The ratio Fe/(Se+Te) in the samples was deviating less than 2~\% from the stoichiometric value indicating a low content of excess iron. The results of the refinement for different samples are given in Table 1. The refined occupation factors for Te and Se ions for the main tetragonal phase are in general agreement with the EDX data. The occupation of iron on the  2c sites was $\sim 6-9$~\%. The values of the lattice parameters for different samples varied within the range 3.800 -- 3.803~{\AA} for the \emph{a(b)} lattice constant, and 6.026 -- 6.047~{\AA} for the \emph{c} parameter and are close to those reported in Ref. 13 for single crystals of similar composition.

\begin{table*}[htb]
\caption{Structural data obtained from the Rietveld refinement for FeSe$_{0.5}$Te$_{0.5}$ samples}\label{tab1}
\begin{tabular}{|l|c|c|c|c|c|c|c|c|c|c|c|}
\hline
Sample & \multicolumn{4}{c|}{Occupation} & \multicolumn{2}{c|}{Lattice constant} & Tetragonal &  Hexagonal & R$_{wp}$ & R$_{exp}$ & $Chi^2$ \\ \cline{2-7}
& Fe1   & Fe2   & Se & Te & \emph{a, b} & \emph{c} & phase & phase &&&\\
& /2a/ & /2c/ & /2c/ & /2c/ & [{\AA}] & [{\AA}] & [\%] & [\%]  & & & \\ \hline
Br N5 & 0.907(3)& 0.093(3)& 0.48(1) & 0.52(1) & 3.8020(4) & 6.0489(9) & 94.5 & 5.5 & 3.53 & 2.48 & 2.02 \\ \hline
F213  & 0.930(5) & 0.070(5) & 0.49(1) & 0.51(1) & 3.8011(2) & 6.0409(7)&  94.5 & 5.5 & 3.68 & 2.54 & 2.10 \\ \hline
F216 step 1 & 0.929(3) & 0.071(3) & 0.49(1) & 0.51(1) & 3.8025(3) & 6.0300(9) & 98.6 & 1.4 & 3.0 & 2.36 & 1.62 \\ \hline
F216 step 2 & 0.937(3) & 0.063(3) & 0.52(1) & 0.48(1) & 3.8000(3) & 6.0257(8)&  97.9 & 2.1 & 3.06 & 2.49 & 1.51 \\ \hline
\end{tabular}
\end{table*}

\subsection{Susceptibility and magnetic hysteresis}

Fig. 3a shows the temperature dependence of the magnetic susceptibility $\chi$ for samples from different growth runs measured on cooling in a field of 10~kOe applied parallel
to the \emph{c}-axis. The samples prepared from the non-purified Se and Te and handled in air have a relatively high susceptibility, $\sim (1-2) \times 10^{-2}$~emu/mol. Below 300~K the
susceptibility continuously increases on decreasing temperature, then shows a pronounced downturn at 125~K and a clear anomaly  at the superconducting
transition at $T_c \sim  14$~K. The downturn at 125~K is related to the Verwey transition of Fe$_3$O$_4$ \cite{V39}, which was also evidenced by electron-spin resonance \cite{ESR} and specific-heat
measurements (see below). The ESR experiment did not reveal any intrinsic absorption of localized moments.

 Samples prepared from the purified elements and handled in the argon box exhibit susceptibility by one order of magnitude lower and do not show
any anomaly at 125~K. The susceptibility of the purest oxygen-free sample, F216 step 1, prepared by fast cooling (1\dgr\,C/min) manifests a non-monotonous temperature behavior
with a broad maximum at around 180~K. It also shows the sharpest anomaly at around 14~K at the superconducting transition. Surprisingly, repeated melting of this sample
followed by slow cooling with the rate of 1\dgr\,C/h, drastically changes the magnetic behavior. The susceptibility of this sample is temperature independent at high
temperatures and shows a paramagnetic tail at low temperatures which might be associated with localized iron ions. Recent investigations of FeSe(Te)
samples with iron excess \cite{LCH09,HBW09,LKQ09} established a localized magnetic iron moment at the 2c sites which can suppress superconductivity. However, both \emph{well} and \emph{poorly} superconducting samples contain comparable amount of the Fe ions at the 2c sites (Table 1). An explanation of the low-temperature susceptibility tail in the poorly superconducting samples due to the hexagonal-phase can be  excluded because the amount of the hexagonal phase can be strongly reduced in the oxygen-free samples. The apparent effect of the hexagonal impurity on the susceptibility can be revealed by comparing the oxygen-free samples F213 and F216 step 1, containing respectively 5.5 and 1.4 \% of the hexagonal phase.

\begin{figure}[htb]
\includegraphics[angle=0,width=0.5\textwidth]{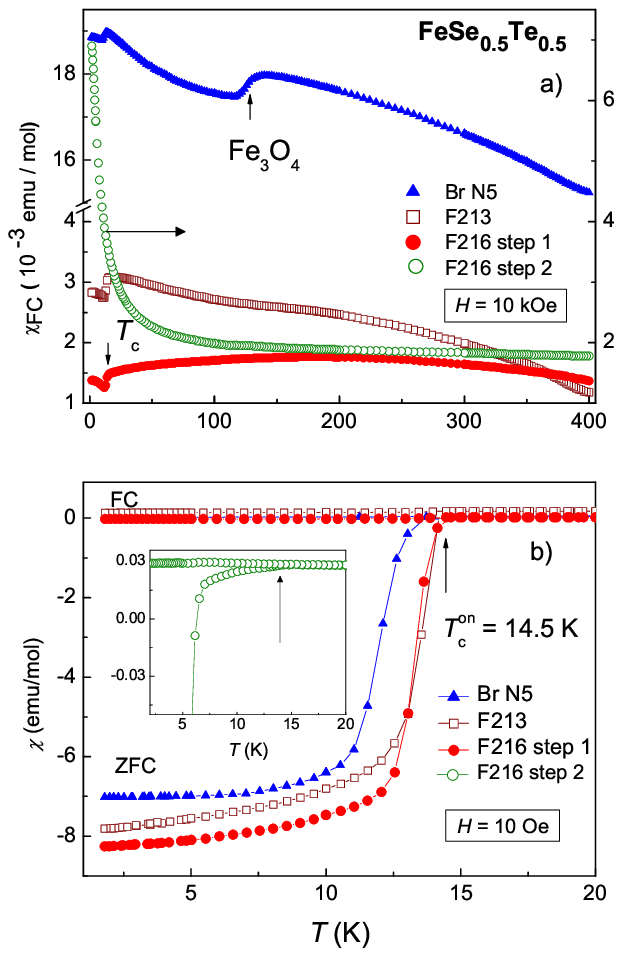}
 \caption{(color online) a) Temperature dependence of the field-cooled susceptibility of FeSe$_{0.5}$Te$_{0.5}$ measured
 in a field of 10~kOe applied along the \emph{c}-axis. The arrow at 125~K
marks the anomaly related to Fe$_3$O$_4$ ; b) Temperature dependences of ZFC and FC susceptibilities of samples with bulk superconductivity
 measured in a field of 10~Oe applied along the \emph{c}-axis.
Arrow indicates the temperature of the onset of the superconducting transition $T_c^{on} = 14.5$~K for the purest samples. Inset: ZFC and FC susceptibilities of the sample with filamentary superconductivity shown on enlarged scale.}
\end{figure}

Fig. 3b shows the temperature dependence of the zero-field cooled (ZFC) and field-cooled (FC) susceptibility measured in a field of 10~Oe applied parallel to the \emph{c}-axis. The measurements along the \emph{ab}-plane showed similar results except that the diamagnetic response was substantially higher for the field parallel to the \emph{c}-axis. The samples containing both, oxide and hexagonal-phase impurities manifest rather broad transitions into the superconducting state. At the same time, the temperature of the onset of the superconductivity, $T_c^{on}$, shows only a small change at around 13.5~K for these samples. We found the sharpest transition and, respectively, the highest $T_c$ of 13.8~K (with $T_c^{on}$ of 14.5~K)  for samples grown from the purified elements under fast cooling conditions only. The values of $4 \pi \chi$ at 2~K calculated from the data shown in Fig. 3b are far above unity indicating that the low-field susceptibility of samples with high diamagnetic response is dominated by demagnetizing effects. The volume fraction of the superconducting phase was therefore estimated from $\chi_{ZFC}$ on needle-like crystals for the field applied along the \emph{ab}-plane with the smallest demagnetizing factors and reached 98~\% for the purest samples. At the same time, in the poorly superconducting sample the onset of the transition remains at the same temperature as in well superconducting samples, but it is masked by a temperature independent paramagnetic background and therefore can be evidenced only on high magnification of the data as shown in the inset in Fig. 3b.  The volume fraction of the superconducting phase with $T_c^{on}$ of 14.5~K in this sample is by more then two orders lower than in the well superconducting sample suggesting the filamentary superconductivity.

Fig. 4 presents the magnetization hysteresis loops for different samples measured at 2~K with the field applied along the \emph{c}-axis. The hysteresis loops for samples with a high volume of the superconducting phase have a symmetric character which indicates dominant bulk pinning and small contribution of surface pinning \cite{SCW09}. Below the transition temperature the diamagnetic response in these samples dominates over the full measured field range. The samples containing oxide impurities show an enhanced (by $\sim 20$~\%) magnetization compared to the oxygen-free samples due to additional pinning centers of Fe$_3$O$_4$.
A strongly contrasting hysteretic behavior was observed for the oxygen-free samples prepared under slow-cooling conditions.
These samples exhibit a considerably reduced width of the hysteresis loops, by two orders of magnitude, suggesting full suppression of the bulk superconductivity.

Surprisingly, such a drastic change of the magnetic properties of the sample with suppressed bulk superconductivity is not
 accompanied by notable changes in the structural data. The x-ray diffraction patterns of these samples show a comparable amount of the tetragonal and impurity phases (Fig. 2a). Their refinement data (Table 1) do not reveal any essential variations of the site occupancy for the iron ions and for the anions as well as of the lattice parameters of the main tetragonal phase. Although any comparison of the low-temperature superconducting data and the x-ray diffraction data taken at room temperature is not plausible, our results indicate a rather subtle role of the structure which probably cannot be  resolved using conventional x-ray facilities.

\begin{figure}[htb]
\includegraphics[angle=0,width=0.5\textwidth]{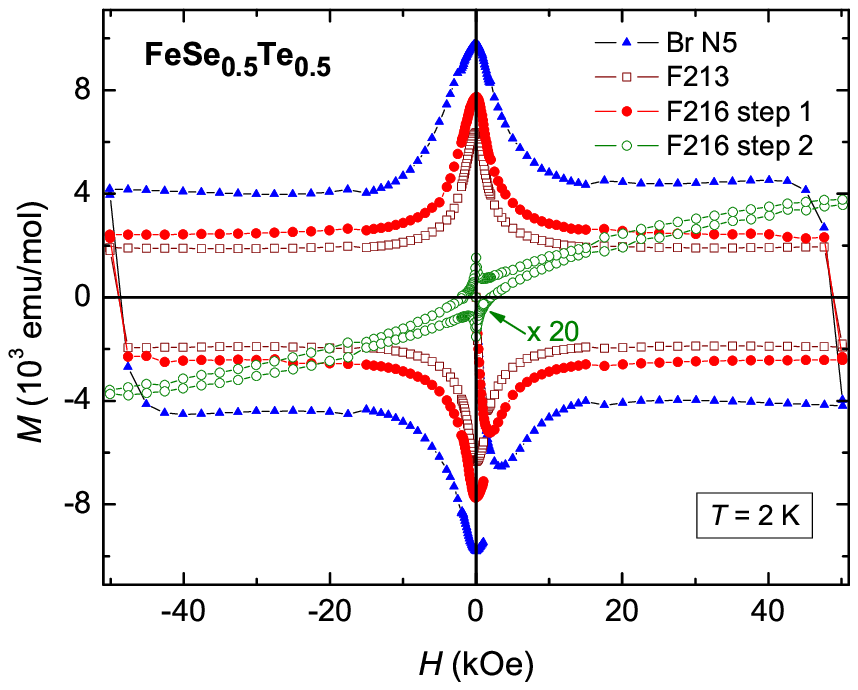}
 \caption{(color online) Hysteresis loops measured at 2~K with the magnetic field applied along the \emph{c}-axis for different single crystalline FeSe$_{0.5}$Te$_{0.5}$ samples. The data for the slowly
cooled sample are magnified by a factor of 20 for clarity.}
\end{figure}

Figures 5a and b show, respectively, the magnetic hysteresis and critical current at different temperatures for the purest samples with bulk superconductivity. The critical current density was estimated from the width of the hysteresis loops using the Bean model for hard superconductors \cite{B62,B64}. At 2 K the  critical current density $j_c$ at zero field reaches a value of $8.6 \times 10^4$~A/cm$^2$.
 For the  sample grown by Bridgman method we obtained $j_c = 9.4 \times 10^4$~A/cm$^2$. For the sample grown from self-flux and intentionally oxidized after the first step we found nearly 3 times higher values of $j_c = 2.3 \times 10^5$~A/cm$^2$. Thus, the oxidation allows to increase the critical current.
 The critical current for the purest sample with high volume fraction of the superconducting phase decreases approximately three times  in fields up to 20~kOe but then flattens at this level up to the largest measured fields suggesting a high current-carrying ability of the material. An estimation of critical currents for $T = 0$ from fits to the experimental data was performed using a generalized power-law dependence $j(T) = j(0)[1-(T/T_c)^p$]$^n$, with $p = 0.5$, $n = 1.5$ and $T_c = 13.8$~K.
  The respective fit is shown by the dotted line in the inset of Fig.~5b, yielding a value $j(0) = 1.7 \times 10^5$~A/cm$^2$.
  The calculated value of $j(0)$
   is close to those determined for high-quality superconducting single crystals of Ba(Fe$_{1-x}$Co$_x$)$_2$As$_2$ \cite{PNT09}.
    Table 2  summarizes the critical current densities calculated from the hysteresis loops at 2~K together with the critical temperature
     $T_c$ and lower critical field $H_{c1}$ determined from the magnetic data.

\begin{table*}[htb]
\caption{Superconducting parameters: transition temperature, critical current, upper and lower
 critical fields for  FeSe$_{0.5}$Te$_{0.5}$ samples}\label{tab2}
\begin{tabular}{|l|c|c|c|c|c|c|}
\hline
Sample & $T_c^{on}$ & $j_c$ & $j_c$ & $H_{c1}$ & $H_{c2}$ & $H_{c2}$ \\
  & (K) & (kA/cm$^2$) & (kA/cm$^2$) & (kOe) & (kOe) & (kOe)
\\ & & (2 K) & (0 K) & (2 K) & [$H \parallel c$] & [$H \parallel ab$]\\ \hline
Br N5 & 13.5 & 94 & 250 & 3 & 510/1300$^*$ & 850\\ \hline
F213  & 14.5 & 77 & & 1.4 & 570/1300$^*$ &\\ \hline
F216 step 1 & 14.5 & 86 & 170 & 1.2 & 490/1500$^*$ &\\ \hline
F216 step 2 & 7.6 & 0.8 & & 0.1 & &\\ \hline
\end{tabular}

$^*$ Estimated from the heat capacity measurements.
\end{table*}

\begin{figure}[htb]
\includegraphics[angle=0,width=0.5\textwidth]{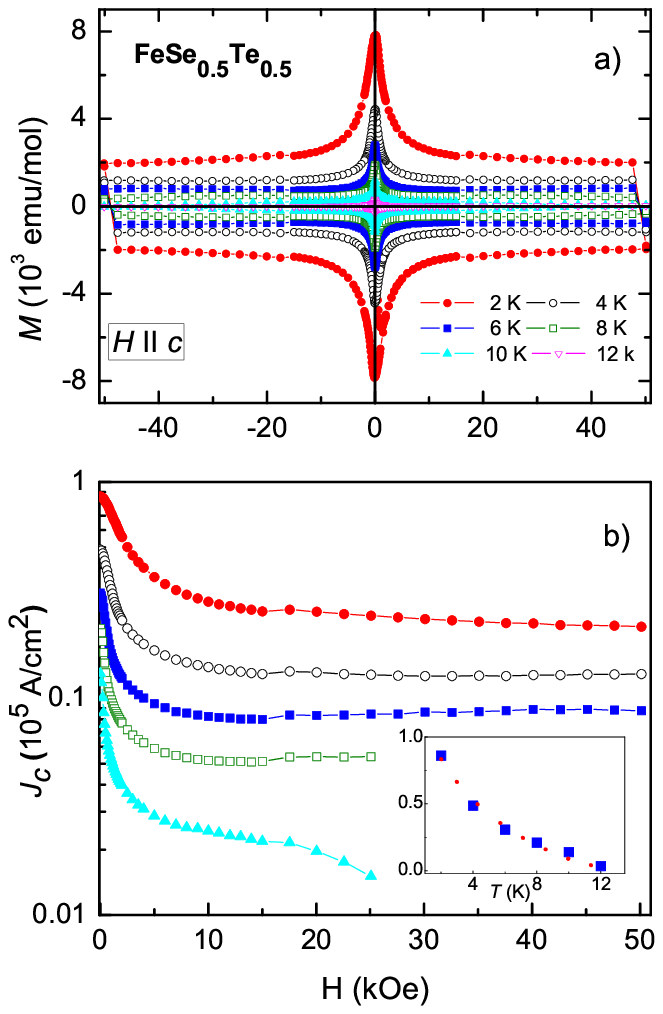}
 \caption{(color online) a) Hysteresis loops at different temperatures measured for the field applied along the \emph{c}-axis for FeSe$_{0.5}$Te$_{0.5}$ sample F216 step 1 grown by self-flux method; b) Critical current density $j_c$ vs. magnetic field at different temperatures for the same sample. The inset shows the temperature dependence of the critical current at zero field with the fit by the power law (dashed line).}
\end{figure}

\subsection{Electrical resistivity}

Fig. 6a presents the temperature dependences of the resistivity for the studied samples.
 The oxygen-free samples have a lower value of the resistivity in the normal state compared to the samples with oxide impurity which can be attributed to increased scattering of charge carriers on additional impurity centers related to Fe$_3$O$_4$. The resistivity of samples with high volume fraction of the superconducting phase exhibits a metal-like temperature dependence below 200~K down to $T_c$ as shown in the inset of this figure. Such a behavior was established earlier only for samples with a low amount of excess iron \cite{LKQ09}.  In contrast, the resistivity of samples with suppressed bulk superconductivity shows a temperature independent behavior at high temperatures down to $T_c$. The resistivity of these samples drops at approximately the same temperature as in the samples with high superconducting parameters, but the resistive transition is strongly broadened and shifted to lower temperatures in agreement with the susceptibility data.

\begin{figure}[htb]
\includegraphics[angle=0,width=0.5\textwidth]{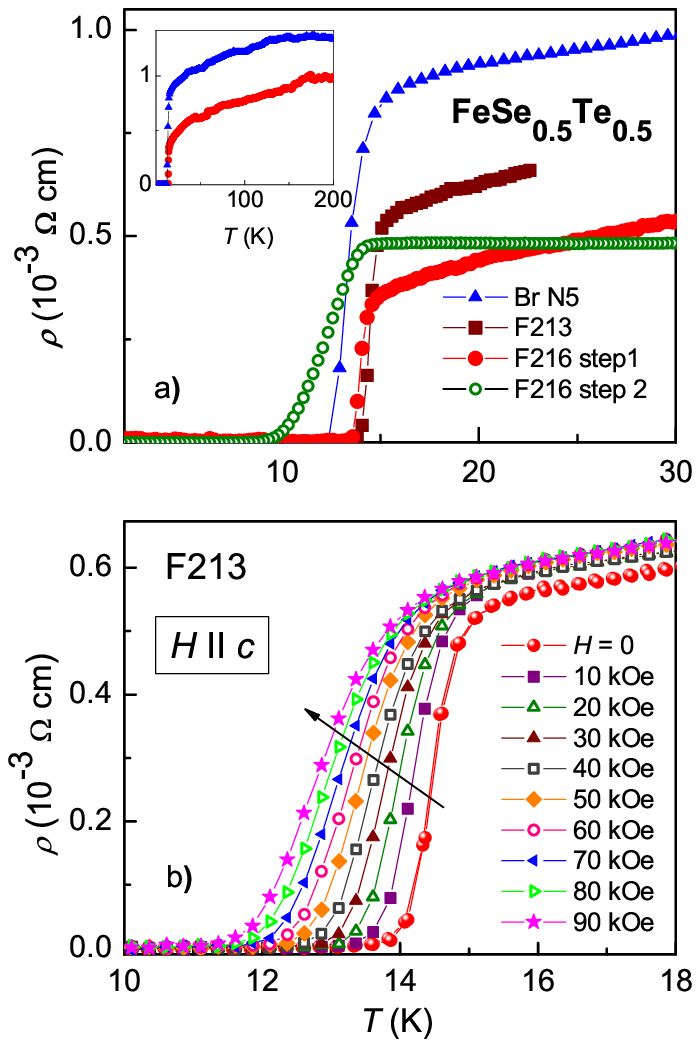}
 \caption{(color online) a) Temperature dependences of the resistivity for different FeSe$_{0.5}$Te$_{0.5}$ samples measured on cooling in zero external magnetic field. Inset: Temperature dependences of the resistivity for samples BrN5 and F216 step 1 measured up to 200 K; b) Temperature dependences of the in-plane resistivity for oxygen-free sample F213 measured at various magnetic fields applied parallel to the \emph{c}-axis. The arrow shows the direction of increasing field.}
\end{figure}

Fig. 6b illustrates the effect of magnetic field on the in-plane resistivity in the transition region for
 one of the oxygen-free samples (F213). The field was applied parallel and perpendicular to the \emph{c}-axis,
 and the measurements were done on warming after cooling in zero field. The resistivity curves are displaced to
 lower temperatures with increasing magnetic field with a stronger shift of the transition for the field parallel to the \emph{c}-axis than for the perpendicular configuration. The obtained data are in general agreement with the respective results on the superconducting FeSe$_{1-x}$Te$_x$ samples with different $x$ reported earlier \cite{HBW09,YMK09}.

In Fig. 7 the temperature dependences of the upper critical field $H_{c2}(T)$ estimated using the criterion of 50~\% drop
 of the normal state resistivity $\rho_n$ are shown. The calculated data show a similar behavior for different samples with
 the shift on the temperature scale corresponding to a difference in their transition temperatures.
On approaching $T_c$ the slope of the $H_{c2}(T)$ curve for the configuration $H \parallel c$ becomes smaller
compared to that of the lower temperatures. Contrary, the slope of $H_{c2}(T)$
 for the configuration $H \perp c$ shows a slight increase on approaching $T_c$. The anisotropy of
 the upper critical field defined as $\gamma_{H_{c2}} = H_{c2}^{ab}/H_{c2}^{c}$ shows a notable
 increase on approaching $T_c$ from $\gamma_{H_{c2}} =2.15$ at $T/T_c = 0.91$ to $\gamma_{H_{c2}} = 3.6$ at $T/T_c = 0.968$
 and finally reaches $\gamma_{H_{c2}} = 6$ at $T/T_c = 0.996$ (see inset of Fig. 7). This observation is in disagreement
 with the nearly isotropic behavior of the upper critical field reported by other studies on superconducting FeSe(S)$_{1-x}$Te$_x$
 with different levels of substitution \cite{HBW09,YMK09}. The higher anisotropy value of $\gamma_{H_{c2}}$
 may reflect the better quality of our samples compared to earlier reports. An estimation of the upper critical field $H_{c2}(0)$ for $T=0$~K
 was achieved within the Werthamer-Helfand-Hohenberg (WHH) model for conventional superconductors, defined by $H_{c2}(0)= - 0.69 T_c (dH_{c2}(T)/dT)\mid_{T_c}$
 for a weak-coupling regime \cite{WHH66}. The calculated data are presented in Table 2. The values of $\mu_0 H_{c2}$(0) are
 rather similar for all samples, are independent of the impurity content, and vary in the range 490 -- 570~kOe for configuration of
 the magnetic field parallel to the \emph{c}-axis. The estimated value of $\mu_0 H_{c2}$(0) for one of the measured samples
  along the \emph{ab} plane
equals $\sim 850$~kOe. Such high values of $H_{c2}(0)$, which are far above the Pauli paramagnetic limit $\mu_0 H_p = 1.84 T_c \sim 260$~kOe can be
attributed to an enhanced impurity scattering from the
Fe at 2c sites as proposed in Ref. 14.

\begin{figure}[htb]
\includegraphics[angle=0,width=0.5\textwidth]{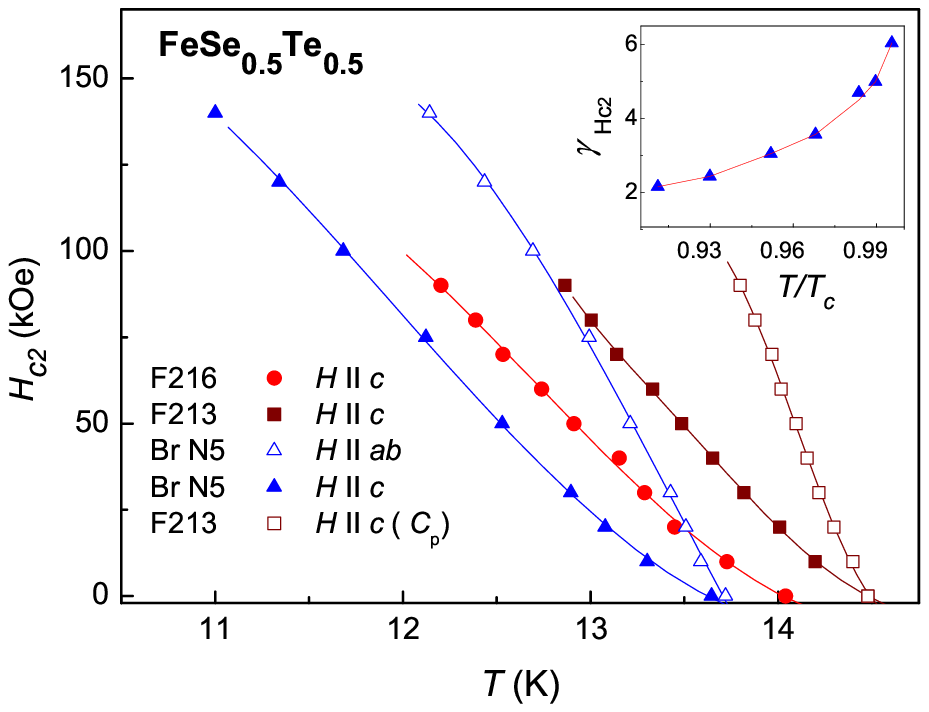}
 \caption{(color online) Temperature dependences of the upper critical field $H_{c2}$ determined using the criterion of $0.5 R_n$ for samples grown by different methods with magnetic field $H$
parallel to the \emph{c}-axis (closed symbols) and parallel to the \emph{ab} plane (open triangles). The open squares represent the data for sample F213 calculated from the specific
heat (see text). The inset shows the temperature dependence of the anisotropy of the upper critical field $\gamma_{H_{c2}} = H_{c2}^{ab}/H_{c2}^{c}$ in the vicinity of $T_c$ for sample Br N5.}
\end{figure}

\subsection{Specific heat}

Fig. 8 shows the temperature dependences of the specific heat for samples with different superconducting properties. They exhibit a similar behavior and similar values of
the specific heat at temperatures above 15~K. A small anomaly at 125~K is observed for sample Br N5 which contains Fe$_3$O$_4$. By scaling the entropy involved in this
anomaly with that of the specific-heat anomaly at the Verwey transition measured in Fe$_3$O$_4$ we estimated the amount of the oxide impurity in this sample to be at a level of
$\sim 1$~mol\%. The inset in Fig. 8 illustrates the specific heat in the representation $C/T$ vs. $T$ in the low temperature range. The samples with bulk superconductivity
exhibit a pronounced anomaly at around 14~K. As the transition in the poorly superconducting sample is smeared out, no discernible anomaly in the specific heat could be detected. The data in Fig. 8 indicate that the superconducting contribution to the specific heat is small compared to that of the lattice contribution which dominates the total specific heat.

\begin{figure}[htb]
\includegraphics[angle=0,width=0.5\textwidth]{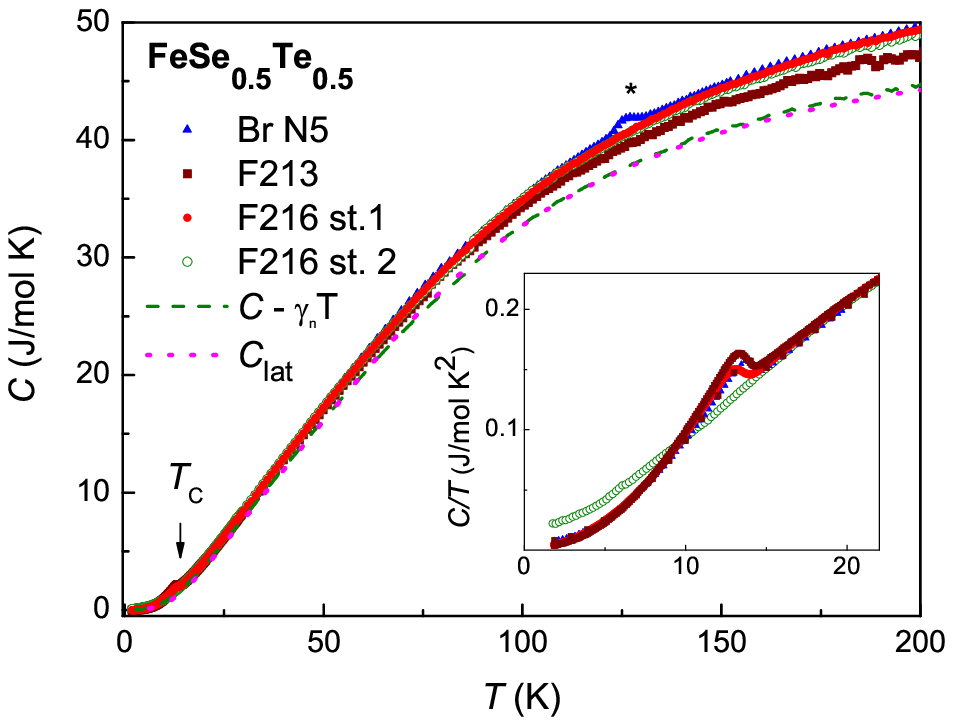}
 \caption{(color online) Temperature dependences of the specific heat for different samples.
  The star indicates the anomaly at the Verwey transition due to presence of Fe$_3$O$_4$ impurity in
sample Br N5. The dashed and dotted lines present the lattice specific heat as described in the text.
 The inset shows the specific heat in the representation $C/T$ vs. $T$ in
the transition region.}
\end{figure}

Fig. 9 shows the temperature dependences of the specific heat plotted as $C/T$ vs. $T^2$ for different samples are shown at temperatures below 5~K. A fit to the experimental data in the range below 4.5~K by using $C/T = \gamma + \beta T^2$ allows to estimate the values of the Sommerfeld coefficient $\gamma$ related to the electronic contribution, and the prefactor $\beta$ which characterizes the lattice contribution to the specific heat in a simple Debye approximation.
 For samples with non-superconducting behavior this procedure gives an estimate of $\gamma$ in the normal state, $\gamma_n$,
  while for superconducting samples it yields the residual $\gamma_r$. The calculated values of these parameters are given
in Table~\ref{tab3}. For the oxygen-free bulky superconducting samples we obtained $\gamma_r = 0.82$ -- 0.96~mJ/mol K$^2$.
The obtained values of the residual $\gamma_r$ for the oxygen-free superconducting samples are much lower than reported
thus far by other authors for similar compositions \cite{SSM09,LKQ09,ZML10}. These extremely low values of $\gamma_r$ confirm the high
purity of our oxygen-free samples. The obtained values of the residual $\gamma_r$ for pure samples indicate that the volume fraction of the superconducting phase reaches $\sim 95$ -- 96\,\% which agrees with the estimate obtained from the susceptibility data. A larger value of $\gamma_r = 5.2$~mJ/mol K$^2$ for sample Br N5 can be probably
 attributed to magnetic contribution from the Fe$_3$O$_4$ impurity although due to insulating nature of Fe$_3$O$_4$ at low temperatures one should not expect any contribution to $\gamma_r$. However, one cannot exclude, for example, the presence of glassy-like disorder in Fe$_3$O$_4$ with low energy excitations linear in temperature or additional magnetic excitations yielding a $T^{3/2}$ dependence of the specific heat.  This problem needs additional study.
 For the samples with suppressed bulk superconductivity we obtained $\gamma_r = 19.3$~mJ/mol~K$^2$.
We must also note that the calculated values of the prefactor $\beta$ for samples with high superconducting parameters
and for those with reduced superconductivity are very close to each other. This indicates that the electronic superconducting
contribution to the specific heat in the range of temperatures used for fitting has only a minor influence
and does not much affect the accuracy of calculations. This justifies the above-mentioned fitting procedure and the correctness
of the obtained parameters.

\begin{figure}[htb]
\includegraphics[angle=0,width=0.5\textwidth]{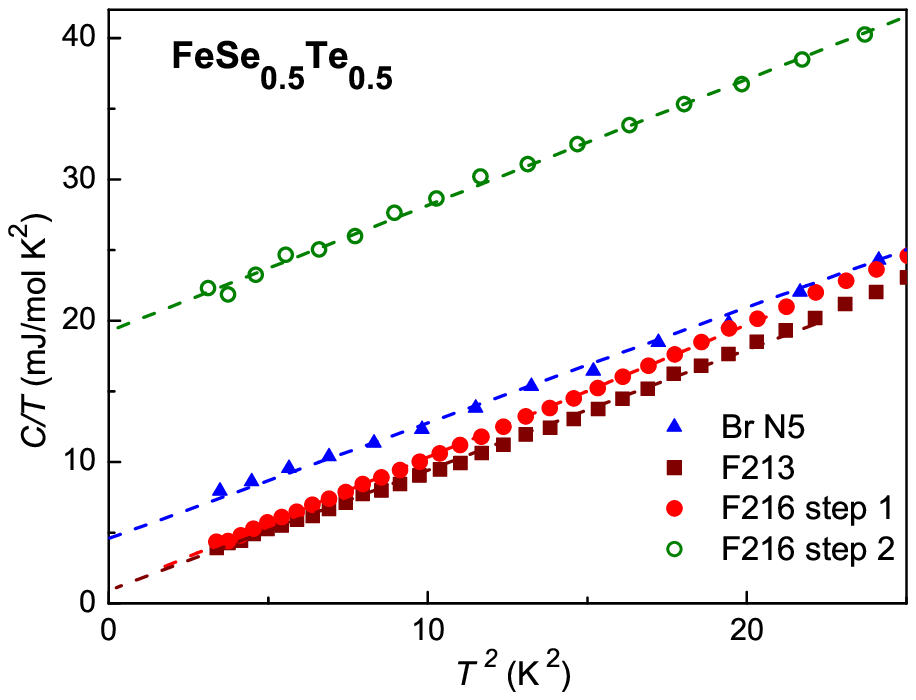}
 \caption{(color online) Temperature dependences of the specific heat in the representation $C/T$ vs. $T^2$ for different samples.}
\end{figure}

To get an additional independent estimate for the electronic and lattice contributions to the specific heat we used the following approach. For a description of the phonon spectrum a combined Einstein-Debye model was employed. The tetragonal unit cell of FeSe(Te) with space group P4/nmm contains two formula units giving rise to a total of 12 normal modes of vibrations. Their contribution was simulated by two Debye terms $C_D$ and one Einstein term $C_E$ with equal distribution of the spectral weight between the Debye and Einstein terms. These assumptions are in rough agreement with the results of the experimental study of the phonon density of states by nuclear
inelastic scattering \cite{KWC10} and neutron scattering \cite{PMT09} on related Fe$_{1+x}$Se superconductors.

The characteristic Debye and Einstein temperatures, $\Theta_D$ and $\Theta_E$, were the input parameters for a fit to the experimental temperature dependence of the total specific heat
above $T_c$ by the expression

\[
C = C_{D1}(\Theta_{D1}) + C_{D2}(\Theta_{D2}) + C_{E}(\Theta_{E}) + \gamma T
\]

The fitting parameters were varied till the minimal deviations from a constant $\gamma_n$ value in a maximal temperature range
 (up to 200~K) for the superconducting samples were
achieved. The temperature dependence of the simulated lattice specific heat with these \emph{optimized} values of
 $\Theta_{D1} = 127$~K, $\Theta_{D2} = 235$~K and $\Theta_E = 315$~K is shown by the dotted line in Fig. 8. For the sample with suppressed bulk superconductivity  by the dashed line Fig. 8 also presents
 a curve calculated by subtracting the normal state electronic contribution $\gamma_nT$ (with $\gamma_n = 23$~mJ/mol~K$^2$)
 from the measured total specific heat. At temperatures above 30~K both curves nicely coincide with deviations less
than 2\,\% in the complete temperature range up to 300~K. This consistency of the data again justifies the model used for
the simulation of the phononic contribution. Note that no scaling of the phononic contribution for samples with high and
 low superconducting parameters was necessary. We found that the estimated values of the
Sommerfeld coefficient in the normal state $\gamma_n$ vary in the range 23 -- 26~mJ/mol~K$^2$ for different samples and are much
lower than those reported previously for FeSe$_{1-x}$Te$_x$ by other
authors \cite{SSM09,LKQ09,ZML10}. To our opinion, the reason of this discrepancy is related to different estimates of
the lattice contribution \cite{EPC1}.

The electronic specific heat $C_e$ for all samples was determined by subtracting the calculated lattice contribution
from the total measured specific heat. The dependences
of the electronic specific heat in the representation $C_e/T$ vs. $T$ are shown in Fig. 10 for different samples
in a temperature range around the transition temperature. All
samples with bulk superconductivity exhibit a sharp anomaly in $C_e$ at $T_c$. The magnitude of the anomaly at $T_c$
correlates with the amount of the superconducting phase in
these samples. In the sample with suppressed superconducting properties no sharp anomaly at $T_c$ is evidenced, but instead
a broad cusp in $C_e$ centered at around 10~K
develops. A magnetic field of 90~kOe fully suppresses the residual filamentary superconductivity in this sample, as show the susceptibility
measurements, but has a negligible effect on the specific heat. This suggests that the broad cusp in the electronic
specific heat is not related to the
superconducting behavior. Importantly, a part of this cusp-like anomaly is also evidenced in $C_e$ for samples with
 bulk superconductivity as a broad right wing at
temperatures above $T_c$. To check whether the broad cusp anomaly can be an artifact due to a possible incorrect description of the lattice specific heat at low
temperatures, the contribution of the lowest phonon was varied up to the limiting value when it coincides with the total specific heat. However, even with this
overestimated lattice contribution, the broad cusp in the electronic specific heat still remains present for all samples, although with somewhat reduced amplitude
(by $\sim 20$\,\%) and with slight shift (by $\sim 1$~K) to lower temperatures. These results indicate that the broad anomaly in the electronic specific heat is not an artifact due to
an inadequate modeling of the phonon density of states, but is a feature reflecting the intrinsic properties of the studied samples. The typical Schottky-like
appearance of this anomaly suggests an electronic origin, and the independence of magnetic field indicates its relation to the orbital degree of freedom. We speculate that
it could originate from a splitting of the ground state of Fe$^{2+}$ ions either by a crystal field or due to spin-orbital coupling. Therefore this anomaly was simulated
within a simple model of a two-level system. The results of the calculation are shown by a solid line for the sample with suppressed superconductivity using the data for
a field of 90~kOe. We arrived at a reasonable description of the cusp at temperatures above 7~K with the value of the ground state splitting $\Delta_{CF} = 24$~cm$^{-1}$ and the amount
of magnetic species corresponding to $\sim 7$\,\% per mole. The only value that correlates with this quantity in our samples is the concentration of Fe ions on the 2c sites as resulting from the refinement of the x-ray data. As already mentioned previously, these 2c site Fe ions can have a local magnetic moment. Therefore it
is natural to assume that they can also be  responsible for the broad cusp-like anomaly in the electronic specific heat. However, the presence of this feature in the
electronic specific heat in the samples with bulk superconductivity just above $T_c$ indicates that the magnetic moment of these Fe ions is not the main factor that destroys superconductivity. A remarkably sharp behavior of the specific heat just below $T_c$ in samples with bulk superconductivity along with extremely low values
of their residual $\gamma_r$ indicates that this contribution is fully suppressed in the superconducting state.

\begin{figure}[htb]
\includegraphics[angle=0,width=0.5\textwidth]{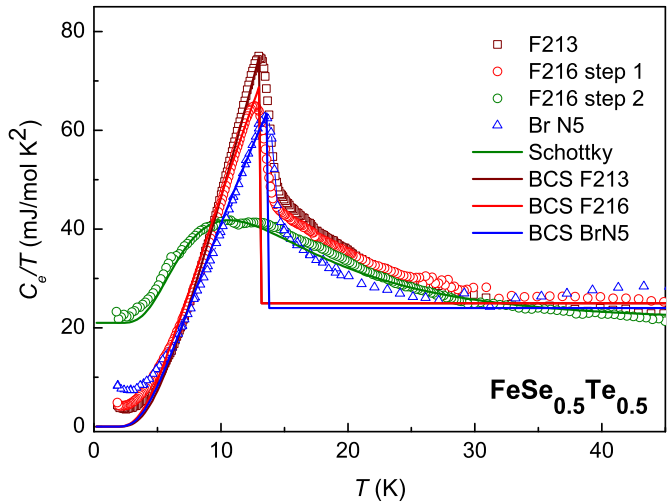}
 \caption{(color online) Temperature dependences of the electronic specific heat in representation $C_e / T$ for different samples.
 The solid lines represent the fits describing,
respectively,  the superconducting specific heat for the bulk superconducting samples within the
single-gap BCS model as given in the text, and the Schottky anomaly in the non-superconducting sample (F216 step 2).}
\end{figure}

\begin{table*}[htb]
\caption{Parameters determined from the heat capacity measurements}\label{tab3}
\begin{tabular}{|l|c|c|c|c|c|}
\hline
Sample & $\gamma_r$ & $\beta$ & $\gamma_n$ & $\Delta_0$ & $2 \Delta_0 / T_c$ \\ & (mJ/mol K$^2$) &  (mJ/mol K$^4$) & (mJ/mol K$^2$) &  (K) &\\ \hline
Br N5 & 5.2 & 0.75 & 24 & 26.6 & 3.94\\ \hline
F213 & 0.82 & 0.85 & 25 & 28.1 & 3.86\\ \hline
F216 step 1 &  0.96 & 0.94 & 25 &  25.9 & 3.57\\ \hline
F216 step 2 &  19.3 &  0.90 & 23 & &\\ \hline
\end{tabular}
\end{table*}

The temperature dependence of the electronic superconducting specific heat was analyzed within a BCS derived $\alpha$-model \cite{BWF01,PNS73}
 with a temperature dependent superconducting gap
 $\Delta$, using a similar approach as described in Ref. 36 used for analysis of the specific heat in related Ba(K)Fe$_2$As$_2$ pnictides.
 In Fig. 10 fitting curves are shown  by
solid lines for three samples with high superconducting parameters. The fitting curves reasonably describe the superconducting specific heat,
 except in the range below 5~K which can be related to effects of residual impurities. The values of the superconducting gap at 0~K, $\Delta_0$, derived from the analysis, vary within the range 26 -- 28~K for different samples. The value of the coupling constant $2 \Delta_0 / T_c$ is close to the BCS value of 3.53.
\begin{sloppypar}
We note that the obtained gap values $\Delta_0 =2.4(1)$~meV are in
good agreement with the value of 2.5~meV derived by Kato \emph{et al.} \cite{KMN09} from scanning tunneling spectroscopy of
Fe$_{1.05}$Se$_{0.15}$Te$_{0.85}$, by Homes \emph{et al.} \cite{HAW10} from optical conductivity of FeSe$_{0.45}$Te$_{0.55}$,
and with 2.6~meV obtained by Biswas \emph{et al.} \cite{BBP10} from $\mu$SR, and by Bendele \emph{et al.} \cite{BWP10} from magnetic penetration studies of FeSe$_{0.5}$Te$_{0.5}$. These studies and
ARPES \cite{CZZ10} indicate a multigap structure of FeSe$_{1-x}$Te$_x$. A fit to the electronic specific heat using a
two-gap model, however, did not give an essential improvement compared to
the single-gap model and therefore we conclude that the specific heat is dominated by only one gap of 2.4~meV.
\end{sloppypar}

\begin{figure}[htb]
\includegraphics[angle=0,width=0.5\textwidth]{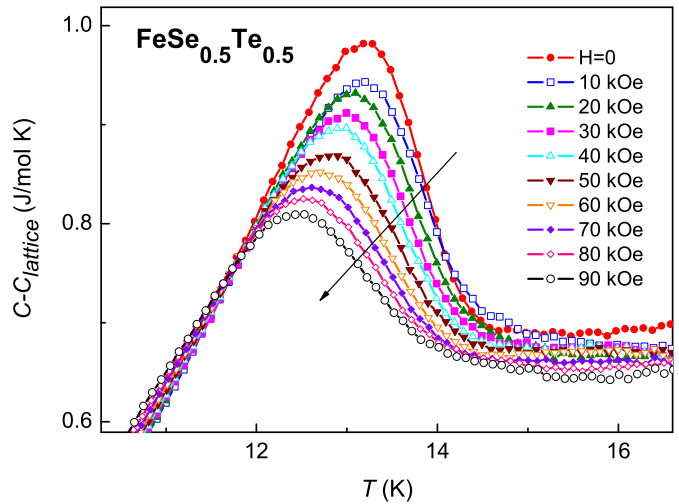}
 \caption{(color online) Temperature dependences of the electronic specific heat at different applied
 magnetic fields for oxygen-free sample F213 with bulk superconductivity. The arrow
shows the direction of increasing field.}
\end{figure}

Finally, in Fig. 11 we present the data for variation of the superconducting specific heat under applied magnetic fields
 for one of the purest samples with high
superconducting parameters. For better presentation the data are shown after subtracting the lattice
contribution from the total specific heat. These dependences allow
to get an estimate of the upper critical field $H_{c2}$ from the temperature shift of the specific heat
 using the criterion of the minimum of the temperature derivative of $C_e$ in
the transition region. The respective dependence $H_{c2} = f(H)$ obtained from these data is shown
by open squares in Fig. 7. Interestingly, it exhibits a behavior resembling
that of the $H_{c2}$ curve determined from the resistivity data for the field applied along the \emph{ab} plane.
 The calculations by the WHH formula \cite{WHH66} gave a value of $H_{c2}(0) \sim 1300$~kOe which is by a factor of 2.5 larger than that obtained from the resistivity data. Similar results were recently
reported for Ba(K)Fe$_2$As$_2$ pnictides by Popovich \emph{et al.} \cite{PBD10}, showing that $H_{c2}(0)$ derived from the specific heat is by a factor of two higher than that calculated from the resistivity, which was attributed to flux-flow effect due to vortex motion. We also would like  to mention that concerning $H_{c2}$ very similar data determined from specific heat and resistivity were recently reported by Serafin \emph{et al.} \cite{SCG10}, which were interpreted as due to the presence of strong thermal fluctuations in FeSe$_{0.5}$Te$_{0.5}$.

\section{Concluding Remarks}

Our detailed studies of the structural, magnetic and thermodynamic properties of FeSe$_{0.5}$Te$_{0.5}$ reveal several important results:

1. Preparation conditions have a
substantial influence on the sample properties: The purity of the starting materials and handling atmosphere are crucial
 for obtaining high-purity samples. Samples
prepared from the purified elements show superior properties compared to other cases. The oxygen-free samples prepared by fast cooling exhibit the lowest values of the susceptibility in the normal state, the highest transition temperature $T_c^{on}$ of 14.5~K, contain a volume fraction
of the superconducting phase up to 98\%, and exhibit the most pronounced anomaly in the specific heat at $T_c$.
In the oxygen-free samples prepared by slow cooling the bulk superconductivity is suppressed. They exhibit a paramagnetic tail in the susceptibility at low temperatures,  a very small width of the hysteresis loop, and a strongly broadened resistive transition.

2. The magnetic hysteresis measurements revealed high values of the critical current density $j_c$ of $8.6 \times 10^4$~A/cm$^2$
for the purest samples which can be attributed to intrinsic inhomogeneity due to disorder at the anion sites.
 The oxidized samples show an increased $j_c$ up to $2.3 \times 10^5$~A/cm$^2$ due to additional pinning centers of Fe$_3$O$_4$.

3. The upper critical field $H_{c2}$ of $\sim 500$~kOe is estimated from the resistivity study in magnetic fields parallel to the \emph{c}-axis for both pure samples and samples containing oxide impurity. The anisotropy of the upper critical field $\gamma_{H_{c2}} = H_{c2}^{ab}/H_{c2}^{c}$ reaches a value of about 6  at $T/T_c = 0.996$ and is the highest reported to date for these materials.

4. The specific-heat measurements evidenced very low values of the residual Sommerfeld coefficient corresponding up to 96\,\% volume
fraction of the superconducting phase which confirms the high quality of the oxygen-free samples.

5. The temperature dependence of the electronic superconducting specific
heat for samples with bulk superconductivity can be reasonably described within a single-band BCS model with the temperature dependent
gap $\Delta_0$ of value 27(1)~K at $T = 0$~K.
The values of the coupling constant $2 \Delta_0 / T_c$ (see Tab. 3) are rather close to the BCS value of 3.53.

6. The electronic specific heat of samples with suppressed bulk
superconductivity shows a broad cusp-like anomaly which is ascribed to a splitting of the ground state of the Fe$^{2+}$ ions
 at the 2c sites. This
contribution is fully suppressed in the ordered state in samples with bulk superconductivity.

\begin{acknowledgements}

The authors thank Dana Vieweg and Nikola Pascher for experimental support. This research has been supported by the DFG via SPP 1458 and Transregional Collaborative Research Center TRR 80
(Augsburg - Munich).

\end{acknowledgements}


\begin{thebibliography}{99}
\bibitem{ESR} The electron-spin resonance spectra of the samples handled in air show broad features characteristic of the Verwey transition observed in magnetite Fe$_3$O$_4$ which was investigated complimentarily. Below the Verwey transition these features in the spectra fully disappear similar to the magnetite sample. The purest (oxygen-free) samples do not show any noticeable features related to the Verwey transition.
\bibitem{EPC1} In several earlier papers on FeSe(Te) a fit by an odd-power polynomial to the experimental data taken just above $T_c$ was used to separate the electronic and lattice
specific heat. Utilizing a similar fitting procedure for the temperature range 15 -- 21~K we obtained for the normal electronic coefficient value $\gamma_n = 90$~mJ/mol~K$^2$ and for
the prefactor $\beta = 0.3$~mJ/mol~K$^4$ corresponding to a Debye temperature $\Theta_D = 235$~K. However, the simple Debye approximation is known to work well only for temperatures below
$\Theta_D / 50 = 4.7$~K \cite{K71}, which is much lower than the temperature range of fitting. Therefore these values of $\gamma_n$ and $\Theta_D$, are strongly overestimated. The failure of the
harmonic-lattice approximation above $T_c$ was demonstrated recently also for the related superconducting BaFe(Co)$_2$As$_2$ pnictides \cite{HWF10}.
\end{thebibliography}
\end{document}